\documentclass[12pt]{article}

\usepackage{epsfig}
\usepackage{graphicx,psfrag}

\title{Certificate Revocation Paradigms}

\author{Jan Willemson\\
Cybernetica, Estonia\\
jan@{}cyber.ee}

\begin{document}

\maketitle

\section{Introduction}
With the extensive development of applications of public key cryptography, the
need for supporting infrastructure arose. It became clear that a person can not
just use another person's public key obtained in some way, but he also has to
\emph{be sure} that the key and its claimed owner really belong together and
that the connection is not expired. 

One of the oldest and most widespread ideologies to satisfy the demands above is
the PKIX one (Public Key Infrastructure using X.509, we refer intersested reader
to \cite{pkix}). The idea is to introduce a Trusted Third Party (TTP) also known
as \emph{Certification Authority} (CA) whose public key is trusted and who 
signs the
certificate ``the person $A$ and the key $N$ belong together at time $t$, this
connection being valid for the time $\delta t$'' by his secret key. 

One major trouble that can occur is the problem of compromise (or otherwise
losing its validity) of $A$'s public key \emph{before} the time $t+\delta t$.
Although the verifier $B$ could have a trusted certificate and check that the
public key has not yet expired, he still should not trust $A$'s public key. To
be absolutely sure, he has to check that the certificate has not been
\emph{revoked}.
It is clear that along with the information about the issuded certificates, the
CA must also distribute the information concerning the revoked ones among them.
The whole problem is finding a suitable infrastructure for that purpose in order
to make the distribution of the revocation information operative and the
structure itself scalable. In what follows we will consider several systems proposed
and discuss their properties.
 
The interested reader may want to learn more about the philosophy of certificate
revocation, in which case (s)he is referred to general discussion papers by Fox
\&\ LaMacchia \cite{foxlam} and Ron Rivest \cite{rivest}.
 

\section{CRL and its improvements} 

The first idea proposed by the authors of PKIX is to collect the information
about revoked certificates into \emph{Certificate Revocation Lists} (CRLs),
i.e. just signed lists of serial numbers of revoked certificates along with some
additional information (see \cite{pkix} for details). 
The first critique can already be made on the basis of
the rough description of the idea --- as it is publicly known, lists are not very
effective data structures. Closer analysis shows that applying only the plain
CRLs, the costs will be unreasonably high (see \cite{mitre} and \cite{micali}
for exact calculations). For instance, if for US Federal PKI there are 
\begin{itemize}
\item three million users, 
\item each CA serves $30000$ users, 
\item 10\% of the certificates are revoked, 
\item CRLs are sent out bi-weekly,
\item the recipient of the digital signature requests certificate information
for 5 signatures per day, and
\item the communication costs are 2 cents per kilobyte,
\end{itemize}
then the total PKI yearly costs are \$732 Millions, of which \$563 Millions are
due to CRL transmission. (Of course, the share of CRL will be greater when more
certifivates are verified per day. For instance if every user verifies 100
sigantures per day, the corresponding amounts are \$10848 Millions and \$10237
Millions.)

So we conclude that in order to get a system with acceptable costs, we need at
least an improvement of CRLs, if not a totally different idea.

\subsection{Delta-CRLs}

The idea of delta-CRLs is in fact the very first attempt to get the system work
more efficiently made by the originators of PKIX (see again \cite{pkix} for
details). The general point of this construction is to issue a full CRL
(called also a base-CRL) after longer inervals and at the meantime provide only
difference lists between the latest base-CRL and the current moment. Now the
users do not need to download all the information (most of it being completely
useless for them) in order to get the freshest revocations, it is enough to get
the last delta-CRL, which has considerably smaller size and thus decreases the
CRL transmission costs. 

Still delta-CRLs do not solve the problem of bottlenecks caused by limited
bandwidth of the network. Just like in the system without delta-CRLs everybody
rushes to make a query for the full CRL just the minute when it is issued, in
the system with delta-CRLs everybody needs the freshest base-CRL in the same
time. A good study on this problem of peak request rate together with
calculations can be found by Cooper in \cite{cooper}. Also the other modifications of CRLs
are treated here using the same reference.

\subsection{Segmented and Staggered CRLs}

One possible way to reduce the transmission costs of CRLs would be to issue the
CRL in segments. We can immediately see that this approach leads to nowhere: the
peak request rate will not drop, but the average request rate will raise (as now
it is not sure that the client gets the necessary information in the first
segment). So we need either to put some additional information together with the
segments or choose a better solution. (As referred above, \cite{cooper} gives a
detailed discussion).

As one may assume that the request rate for a CRL segment declines over time, it
might be a good idea to stagger the issuance of the segments and to issue e.g.
four segments with intervals of 6 hours instead of issuing them together.
Unfortunately it apears that increasing the number of intervals gives a quick
return by making the request rate high all the time, as shown by Cooper
\cite{cooper}.

\subsection{Over-issued CRLs}

One more possible solution to get the request rate lower is not to wait until
the \texttt{nextUpdate} time of the CRL, but to issue a new CRL earlier. So
all the time there will be several not-yet-expired CRLs. This solution really
works and the peak request rate will drop proportionally to the increase of the
number of the over-issuances per lifetime of the CRL. (See Cooper \cite{cooper}.)

There exists another modification of the over-issued CRLs and delta-CRLs is suggested in 
\cite{cooper}. The idea is to issue a delta-CRL together with every 
over-issued CRL, but having the oldest not-yet-expired CRL as its base-CRL. The
modification also includes the need for changing of semantics of delta-CRL so
that a bit more information is included. Namely the author suggests to include the
history of the certificates between the issuance of this delta-CRL and its
base-CRL, e.g. the information about the certificates that were on hold.


\section{IETF's OCSP}

The whole PKIX ideology (including CRLs) is being developed under IETF. On the
other hand, as IETF is not a fixed organization with all its members thinking
the same way, there are often different approaches to the same
problem. This holds true also for certificate revocation matters. An IETF
working group has developed its own view on certificate revocation called
\emph{Online Certificate Status Protocol} (OCSP). There are essencially two
Internet Drafts on this topic referred here as \cite{ocsp} and
\cite{ocsp-caching}. The first of them does not describe much protocols (as one 
might expect by its title), but mostly gives syntax for the status request and
response. The second draft, on the contrary, deals with details of the
information transmission and caching in the OCSP network, being much more
exhaustive and human-readable.

The general idea behind OCSP is to forget about the inconvenient CRL and
concentrate on the important matters. Indeed, if $B$ wants to know, whether
$A$'s public key is still valid, he is not interested in getting the list of all
the revoked certificates, he would rather prefer to have just one short answer
of ``yes'', ``no'' or at least ``not known''. In order to provide the client
with such an answer, a new TTP (called \emph{OCSP Responder} 
in \cite{ocsp}) is introduced. 

An OCSP Responder is generally a server which is able to receive a request about
the current status of a certificate and do the following:
\begin{itemize}
\item check whether it has the necessary information in its cache, and if it is
not older than the age accepted by the client, send the response;
\item if the information is not cached or is too old, froward the request to one
or more neighbouring OCSP Responders and cache the answer when it goes back to
the client; and
\item if the OCSP Responder under consideration is a very special one, having
direct access to the CA's database and called ``co-located'' with the CA, it can
issue the necessary response itself (and, of course, cache it).
\end{itemize}
The designers of the OCSP consider the last kind of OCSP Responders as forming
one unit together with its CA and so they \emph{together} are called a CA in
the drafts. 

Thus, at the same time when the aim of creating the CRL framework was to get the
revocation information handling away from CA in order to reduce the number of
requests made to CA, OCSP takes the source of responses back to CA. Of course,
the caching is meant to assure that the status of the same certificate is not
requested too often from CA, but is rather answered using a cache further away 
of it. There are no estimates given in the drafts concerning the time consumed
by issuing a CRL or answering the online requests. Still, assuming that some of
the revoked certificates are never queried and thus are included in CRL without
much sense, the time needed to complete the two tasks should be at least
comparable. Besides that, the online orientation of the OCSP much better matches
the online ideology of today on the contrary to the offline considerations which
were important at the birthtime of X.509. The authors of OCSP do not exclude the
possibility of using CRLs e.g. as one source of revocation information for 
the OCSP Responder, but in fact OCSP can manage perfectly without them.


\section{CRS}\label{sec:CRS}

In fact all the certificate revocation paradigms make use of an intermediate
layer between CA and the end-user. These layers were for instance CRL
distribution points for PKIX and OCSP Responders for OCSP. In 1995 Silvio Micali
proposed another approach to this layer called \emph{Certificate Revocation
Status Directory} (CRS Directory or just Directory in Micali's terms, see
\cite{micali}). The idea
behind CRS is to alter the underlying datastructure from the 
raw list of all kind of information to something more compact and in such a 
way to reduce the amount of data transmission required.

The setup of the system begins with producing a conventional certificate
(including the public key, keyholder's distinguished name, date of the
certification, date of expiry etc), but adding 200 more bits of information:
\begin{itemize}
\item 100-bit value $Y$ (standing for ``YES, the certificate is good''), and
\item 100-bit value $N$ (standing for ``NO, the certificate is no more good'').
\end{itemize}
The system's running round is one year and the revocation information is updated
each day, so in the case the certificate is still good, we need a confirmation
for it each day. On that purpose 365 ``YES''-answers are hidden in $Y$ the
following way. First, CA chooses a one-way function 
$F:\{0,1\}^{100}\rightarrow\{0,1\}^{100}$ and makes it public. Next it chooses a
secret random value $Y_0\in\{0,1\}^{100}$ and applies the function $F$ 365 
times to it obtaining $Y=F^{365}(Y_0)$. This $Y$ will be included to the 
certificate proving that the certificate was valid at the 0th day. 

In order to prove that the certificate is still valid at the first day, the CA
will calculate $F^{364}(Y_0)$ and transmit it to the Directory. CA is of course
able to do so, as it knows the secret value $Y_0$ and can apply the function $F$
364 times to it. For every other party the
computation of $F^{364}(Y_0)$ knowing only $F^{365}(Y_0)$ would mean inverting 
the one-way function $F$, which is assumed to be computationally infeasible. 

This procedure will be carried out day-by-day, each time opening the original of
the last day's $Y$-value, until the certificate remains valid. 

When the certificate loses its validity, the CA will need the $N$-value included
in the certificate. It is calculated similarily to the $Y$ by choosing a random
secret $N_0\in\{0,1\}^{100}$ and calculating $N=F(N_0)$. This time we will need
only one original -- if the certificate is revoked one day, it remains revoked
the next day and the same issued ``NO''-value can prove that.

The basic idea of the rest of the system is now easy to understand. Let every
certificate ossued by the CA have 20-bit serial number. Then the set of all
issued and not-yet-expired certificates can be characterized by a $2^{20}$-bit
string, where 1 stands for ``not-yet-expired'' (including the certificates revoked
that day!) and 0 stands for ``expired'' or
``unissued''. First on the day number $i$, CA sends to the directory this 
(signed) string and next for all 1s in the string it sends also the value 
$F^{365-i}(Y_0)$ if the certificate is still valid and the value $N_0$ if it was
revoked that day. (As an additional feature the CA may send to the Directory
also the information concerning the reason of revocation.)

Now getting the query from a user $U$ about the status of a certificate, 
the Directory simply finds this 100-bit value relative to that certificate 
and sends it to $U$. The user can now check whether it is the ``YES''-value by
applying the function $F$ $i$ times, or the ``NO''-value by
applying the function $F$ one time. Note that the values $Y$ and $N$ were
included into the certificate, so there is no need to get them separately.

It is a bit unclear from Micali's article \cite{micali}, how should $U$ know the
right Directory (as there can be several of them) and how should the directory
prove that it does not have any information about a particular certificate. But
these are rather the technical details. Much more important are Micali's results
concerning the cost of his system. According to the calculations given in 
\cite{micali} its cost is approximately 900 times cheaper than the one found in
\cite{mitre} for the basic CRL system. This difference is achieved mainly on
reducing the interaction between Directory and the user. On the other hand, the
communication rate between CA and Directory might increase, as pointed out by
Naor and Nissim \cite{naornissim}. They also propose an improvement over
Micali's scheme using hash trees. We will see similar ideas in the next
sections.


\section{HCRS}

In this section we review the paper \cite{hcrs} by Aiello, Lodha and Ostrovsky.
They start with the observation that in the Micali's scheme the CA sends to the
Directory exactly the same amount of information about each certificate. The
authors try to reduce this by organizing a clever tree and introducing a
notion of \emph{complement cover families}. We will call their system a
\emph{Hierarchical Certificate Revocation Scheme} (HCRS).

\begin{figure}[t]
\begin{center}
\psfrag{f}{$\phi$}
\includegraphics{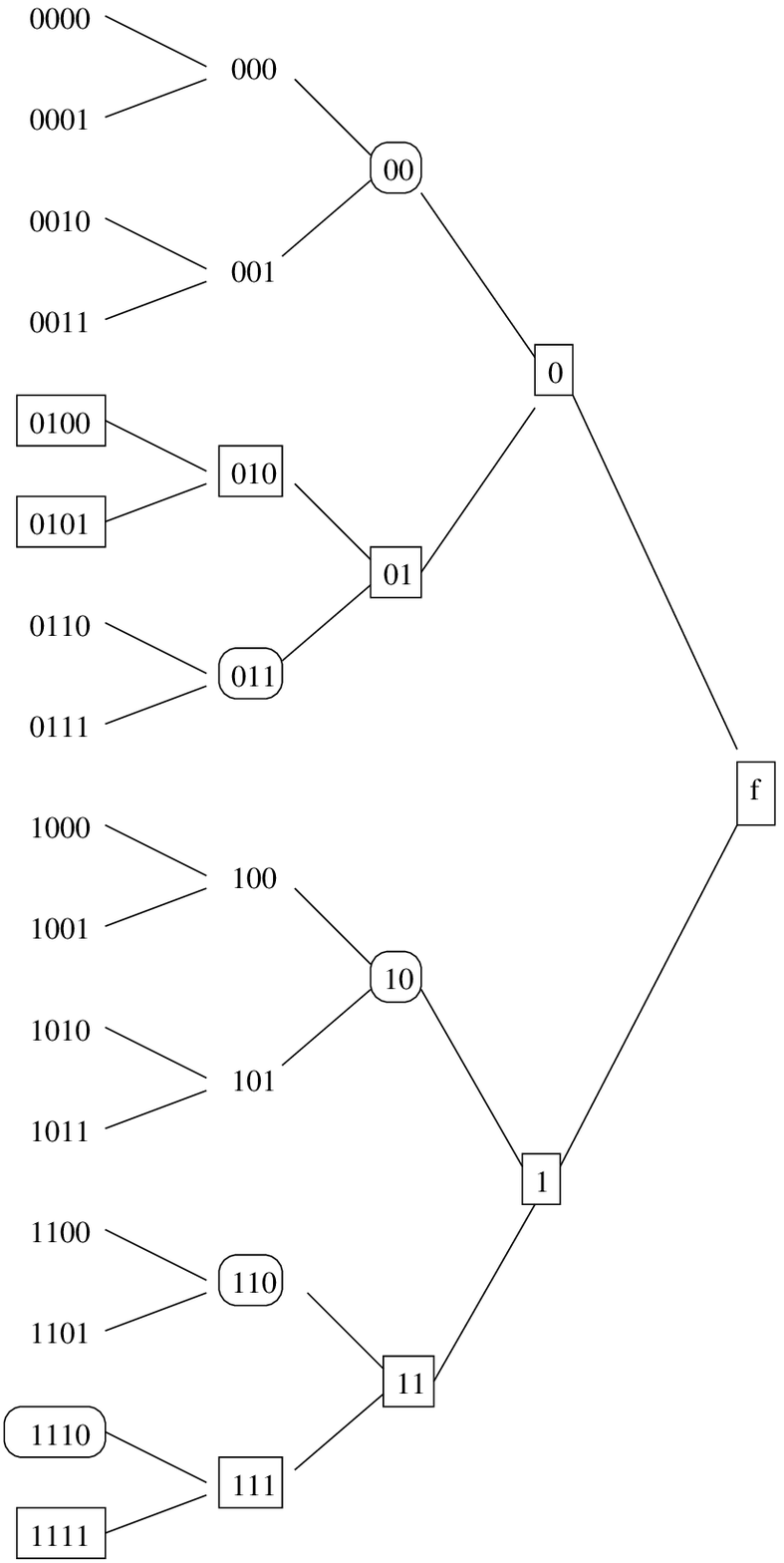}
\end{center}
\caption{Hierarchical Certificate Revocation Scheme}\label{fig:hcrs}
\end{figure}

Let us first for simplicity have $2^l$ certified 
users denoted by binary vectors of length
$l$. Now take $2^{l-1}$ binary vectors of length $l-1$ and make the vector
$u_1\ldots u_{l-1}$ to be the parent for the vectors $u_1\ldots u_{l-1}0$
and $u_1\ldots u_{l-1}1$. Continuing in the same fashion we end up with 
binary tree having the empty vector $\phi$ as its root. (See Figure
\ref{fig:hcrs}.)

Let $F$ be a one-way
hash function and $D$ the validity time for the certificates as in the Micali's
scheme. Now for each node of the tree (both leaves and internal nodes!) the CA
calculates the sequence $(r,F(r),F^2(r),\ldots,F^D(r))$ where $r$ is a
random initial value for that node. This construction resembles to the Micali's
``YES''-value computation, see section \ref{sec:CRS}. But in this system 
the ``YES'' values
$F^D(r)$ of all the nodes on the path from the leaf $v$ to the root $\phi$ are included
in the $v$'s certificate. 

Now let us have the set $R_i$ of users revoked up to day number $i$. At the end
of that day the CA will calculate a subset of the nodes of the binary tree constructed above.
This subset will be called the set of the \emph{day-$i$ verification nodes} and
it must satisfy the following two conditions.
\begin{enumerate}
\item For every $v\not\in R_i$ there is at least one node on the path from $v$
to the root $\phi$ which is not the day-$i$ verification node.
\item For every user $w\in R_i$ none of the nodes on the path from $v$
to the root $\phi$ is a day-$i$ verification node.
\end{enumerate}
In such a way the valid (non-revoked) users are exactly those for whom there
exist a day-$i$ verification ancestor and the revoked users are exactly those 
for whom none of the ancestors is a day-$i$ verification one (we consider $u$ to
be his own ancestor, too). 

As the next step CA sends to the Directory the values $F^{D-i}(r)$ for each 
day-$i$ verification node are sent to the Directory. For the response to the
question about the validity of $u$'s certificate on the day $i$ the Directory
can send any of the values $F^{D-i}(r)$ from the path from $u$ to the root
$\phi$, if $u$ is not revoked. If $u$ is revoked, the Directory is not able to
produce such a value.

Consider an example. Let there be 16 users labelled by binary vectors of length
4 and form the tree as shown in the Figure \ref{fig:hcrs}. Suppose that the
certificates of the users 0100, 0101 and 1111 are revoked. By the definition of
the day-$i$ verification node we see that these nodes automatically exclude the
nodes 010, 111, 01, 11, 0, 1 and $\phi$ (marked by rectangles on the Figure
\ref{fig:hcrs}). A suitable (and in fact, the samllest) set of the day-$i$ verification nodes is for
instance $\{00,011,10,110,1110\}$ (marked by round boxes).

It is easy to see that for every possible set $R_i$ there exists a set of the
day-$i$ verification nodes and it is generally not a unique one. For the sake of
the low transmission costs it would of course be necessary to be sure that there
always exists a suitable set of reasonable size and that this set is easily
computable. The authors show in \cite{hcrs} that if the number of users is
$N(=2^l)$ and the number of the revoked certificates is $R$, then the number of
values to be transmitted is upper bounded by $R\cdot\log(N/R)$. The authors also
present a possible generalization of the binary hash tree to a (modification of)
$c$-ary tree and they obtain the similar upper bound $R\cdot\log_c(N/R)+R$.


\section{CRT}

Paul Kocher \cite{crt} proposed another method based on trees which this time
are hashed trees called \emph{Certficate Revocation Trees} (CRT).

The main idea of this system is building of the tree using the
revocation information obtained from CAs. This will be done by special authority
called the \emph{CRT issuer}.

A CRT issuer can serve several CAs and his basic action is to write down
statements about the numbers of the revoked certificates so that each possible
certificate number (revoked or not, issued or not) matches \emph{exactly} one of
these statements. After this he hashes these statements into a binary hash tree
and publishes the value of the root hash. If a user requests  information
about a certain certificate, the CRT issuer finds the matching statement and
sends it back together with the necessary intermediate hashes required to
calculate the root value.

Let us consider an example due to Paul Kocher, copy'n'pasted from \cite{crt}.

We have three CAs with 
public key hashes $CA_1 < CA_2 <CA_3$, where $CA_1$ has revoked 3 certificates 
(156, 343, and 344), $CA_2$
has revoked no certificates, and $CA_3$ has 1 revoked certificate (987).
The CRT issuer can now make the following statements about
certificate serial number $X$ from a CA whose public key hash is
$CA_X$: 

\begin{tabular}{ll}
   If: $-\infty < CA_X < CA_1$&
                     Then: Unknown CA \\ &(revocation status unknown)\\
   If: $CA_X = CA_1$ and $-\infty \leq X < 156$&
                     Then: $X$ is revoked \\ &if and only if $X = -\infty$.\\
   If: $CA_X = CA_1$ and $156 \leq X < 343$&
                     Then: $X$ is revoked \\ &if and only if $X = 156$.\\
   If: $CA_X = CA_1$ and $343 \leq X < 344$&
                     Then: $X$ is revoked \\ &if and only if $X = 343$.\\
   If: $CA_X = CA_1$ and $344 \leq X < \infty$&
                     Then: $X$ is revoked \\ &if and only if $X = 344$.\\
   If: $CA_1 < CA_X < CA_2$&
                     Then: Unknown CA \\ &(revocation status unknown).\\
   If: $CA_X = CA_2$ and $-\infty \leq X < \infty$&
                     Then: $X$ is revoked \\ &if and only if $X = -\infty$.\\
   If: $CA_2 < CA_X < CA_3$&
                     Then: Unknown CA \\ &(revocation status unknown).\\
   If: $CA_X = CA_3$ and $-\infty \leq X < 987$&
                     Then: $X$ is revoked \\ &if and only if $X = -\infty$.\\
   If: $CA_X = CA_3$ and $987 \leq X < \infty$&
                     Then: $X$ is revoked \\ &if and only if $X = 987$.\\
   If: $CA_3 < CA_X < \infty$&
                     Then: Unknown CA \\ &(revocation status unknown).
\end{tabular}

Note that for any $CA_X$ and $X$ there is a unique appropriate statement
above which provides all known information about whether X is revoked.

\begin{figure}[t]
\begin{center}
\epsfig{file=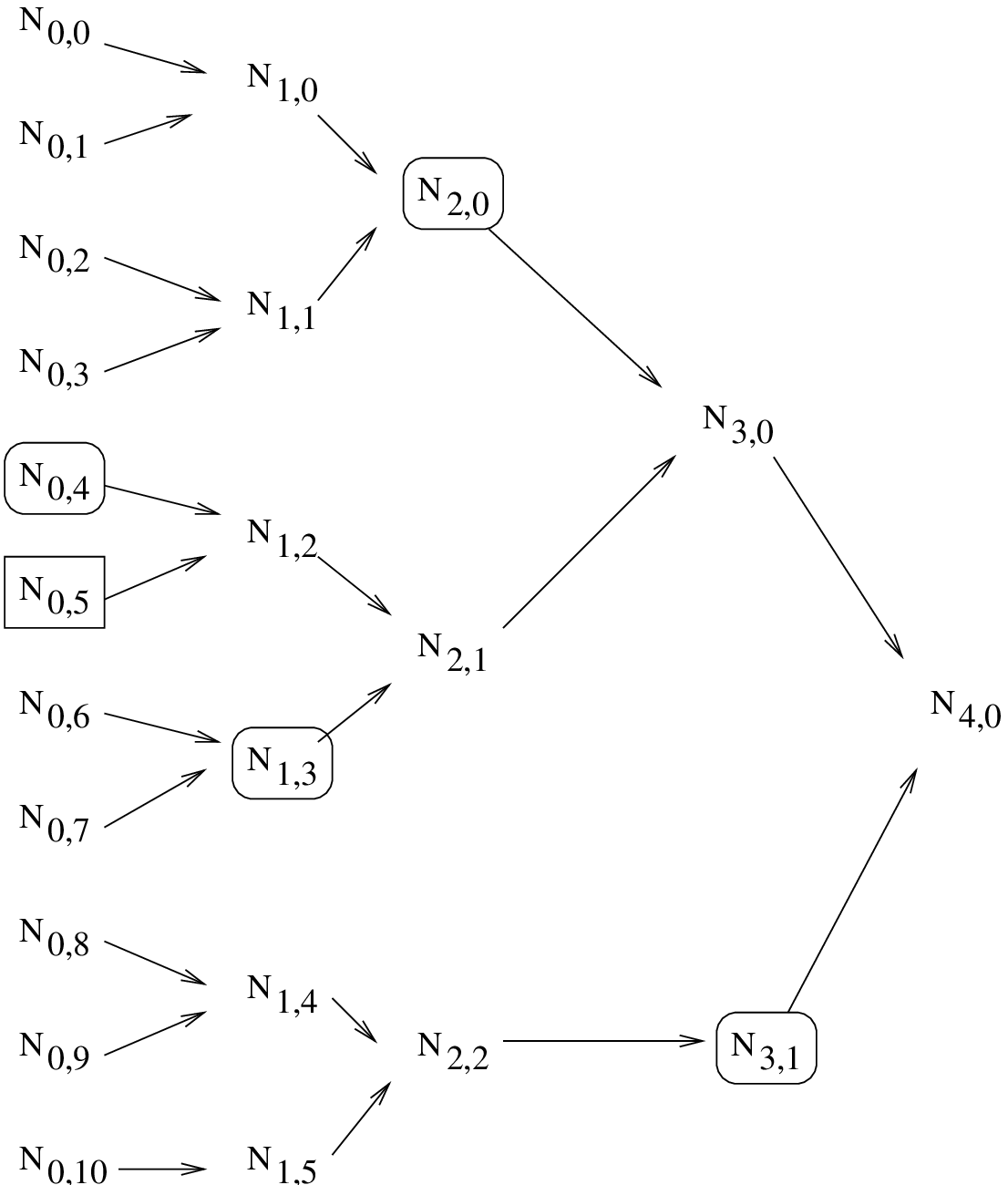}
\end{center}
\caption{Certificate Revocation Tree structure}\label{fig:crt}
\end{figure}

The CRT issuer will now construct a hash tree as follows. Let the statement
``If: $-\infty < CA_X < CA_1$ then: Unknown CA'' (written down using some
standard data structure) be
denoted as $N_{0,0}$ and the other statements from above similarily as 
$N_{0,1},\ldots,N_{0,10}$. Now the tree structure shown in the Figure
\ref{fig:crt} will be established. The figure shoud read like this: the value of
the node $N_{i,j}$ will be computed as the hash of the values in the node(s)
prior to $N_{i,j}$ in the tree. Thus for example $N_{2,2}=H(N_{1,4}|N_{1,5})$,
where $H$ is a hash function and ``$|$'' denotes concatenation.

The root value $N_{4,0}$ will be made public and the proof that the value
$N_{0,i}$ took part in the formation of the tree will be the set of nodes
necessary to calculate the root value. 

For example, if a verifier wants to check the validity of the certificate number
600 from $CA_1$, the appropriate statement he obtains from CRT issuer is: 
``If: $CA_X = CA_1$ and $344 \leq X < \infty$, then: $X$ is revoked if and 
only if $X = 344$.
The verifier can hash this statement structure to get $N_{0,5}$. The
supporting nodes in this example are $N_{0,4}$, $N_{1,3}$, $N_{2,0}$
and $N_{3,1}$ (see the Figure \ref{fig:crt}). The verifier can now use the secure hash function H to compute: 
\begin{eqnarray*}
      N_{1,2}& =& H(N_{0,4} | N_{0,5})\\ 
      N_{2,1} &= &H(N_{1,2} | N_{1,3}) \\ 
      N_{3,0} &= &H(N_{2,0} | N_{2,1}) \\ 
      N_{4,0} &=& H(N_{3,0} | N_{3,1})
\end{eqnarray*}

As the very last remark it should be mentioned that this system is not just a
theoretical view on certificate revocation. It is also implemented and used by
ValiCert company.


\section{Naor's and Nissim's scheme}

Naor and Nissim \cite{naornissim} take Kocher's Certificate Revocation Trees as
a starting point and use a bit more clever data structure than just a(n
incomplete) binary tree. They note that using CRT scheme, the CRT issuer needs
to recompute the full tree at every update. Since there can be only a few
modifications at a time, it seems quite unreasonable. 

The tree used by Naor and Nissim is a \emph{2-3 tree}, i.e. tree where every
interior node has 2 or 3 children and the paths from the root to the 
leaves have the
same length. This structure has the following good properties:
\begin{itemize}
\item testing membership, inserting and deleting a single element are done in
logarithmic time;
\item inserting and deleting of an element affect only the nodes on the
insertion/deletion paths.
\end{itemize}
From the second property we already see the major point of the article --- every
change of the leaf affects only a few interior points, so there is no need for
recalculating the whole tree. Besides, using this system it is not necessary to
invent some kind of additional data structure for the statements like in the
case of CRL there was the structure of statement. Now the tree issuer (Naor and
Nissim consider the tree issuance by CA himself) can put the plain serial
numbers of the revoked certificates into the leaves. In the case of adding the new
revocations or deleting some revoked certificates (e.g. due to expiry) the
issuer (CA) can insert/delete the necessary nodes, recompute the values on the
paths of these nodes, send the results to the Directory and publish the new root
hash value. 

Answering the requests of the users, the Directory sends the values of the
additional nodes required to compute the root value, if the corresponding serial
number can be found in the tree. If not, the Directory finds two neighbouring
leaves, between which the requested certificate should be located and sends the
similar proving information for those leaves (here we
assume that the leaves are ordered, e.g. in the increasing order of serial
numbers). 

Note that due to tree issuance by CA, it is not necessary to have additional
trust in the tree issuer as it was in the case of CRTs. Of course, it is
possible to issue CRTs by CAs as well, in which case we probably lose the
possibility of serving several CAs (not having trust in each other) by one CRT
issuer.



\begin{thebibliography}{99}

\bibitem{pkix}
R. Housley, W. Ford , W. Polk, D. Solo, \textit{Internet X.509 Public Key 
Infrastructure: Certificate and CRL Profile}, Internet Draft\\
\texttt{draft-ietf-pkix-ipki-part1-11.txt}

\bibitem{mitre}
Shimshon Berkovits, Santosh Chokhani, Judith A. Furlong, Jisoo A. Geiter, 
Jonathan C. Guild, \textit{Public Key Infrastructure Study: Final Report} 
Produced by the MITRE Corporation for NIST, April 1994, availiable from\\
\texttt{http://csrc.nist.gov/pki/documents/mitre.ps}

\bibitem{micali}
Silvio Micali, \textit{Efficient Certificate Revocation}, availiable from\\
\texttt{ftp://ftp-pubs.lcs.mit.edu/pub/lcs-pubs/tm.outbox/\\
MIT-LCS-TM-542b.ps.gz}

\bibitem{cooper}
David A. Cooper,  \textit{A model of certificate revocation} (draft), 
February 24, 1999, availiable from\\
\texttt{http://csrc.nist.gov/pki/documents/crlmodel.pdf}

\bibitem{ocsp}
Michael Myers, Rich Ankney, Ambarish Malpani, Slava Galperin,
Carlisle Adams, \textit{X.509 Internet Public Key Infrastructure:
Online Certificate Status Protocol - OCSP}, Internet Draft\\
\texttt{draft-ietf-pkix-ocsp-07.txt}

\bibitem{ocsp-caching}
M. Branchaud, \textit{Internet Public Key Infrastructure:
Caching the Online Certificate Status Protocol}, Internet Draft\\
\texttt{draft-ietf-pkix-ocsp-caching-00.txt}

\bibitem{naornissim}
Moni Naor, Kobbi Nissim, \textit{Certificate Revocation and 
Certificate Update}, 7th USENIX Security Symposium, 1998, availiable from\\
\texttt{http://www.wisdom.weizmann.ac.il/~naor/onpub.html}

\bibitem{hcrs}
William Aiello, Sachin Lodha, Rafail Ostrovsky,
\textit{Fast Digital Identity Revocation} (Extended Abstract), in 
\textit{Advances in Cryptology -- CRYPTO'98}, pp 137-152, Springer 1998

\bibitem{foxlam}
Barbara L. Fox,  Brian A. LaMacchia, \textit{Certificate Revocation: 
Mechanics and Meaning}, availiable at\\
\texttt{http://www.farcaster.com/papers/fc98/index.htm}

\bibitem{rivest}
Ronald R. Rivest, \textit{Can We Eliminate Revocation Lists?}, to appear in
the Proceedings of Financial Cryptography 1998, availiable at\\
\texttt{http://theory.lcs.mit.edu/~rivest/revocation.ps}

\bibitem{crt}
Paul Kocher, \textit{A Quick Introduction to Certificate Revocation Trees},
availiable at\\
\texttt{http://www.valicert.com/technology/body.html}

\end{thebibliography}
\end{document}